\documentclass[conference]{IEEEtran}
\IEEEoverridecommandlockouts
\usepackage{amsmath}
\usepackage{balance}
\usepackage{cite}
\usepackage{acronym}
\usepackage{subfigure}
\usepackage{hyperref}
\usepackage{color}

\usepackage{flushend}

\usepackage{graphicx}
\usepackage{amsmath}
\usepackage{amssymb}
\usepackage{multicol}

\newcommand{\argmax}{\arg\!\max}

\acrodef{SC-FDMA}[SC-FDMA]{single-carrier frequency-domain multiple-access}
\acrodef{LOS}[LOS]{line-of-sight}
\acrodef{NLOS}[NLOS]{non-line-of-sight}
\acrodef{SINR}[SINR]{signal-to-interference-and-noise ratio}
\acrodef{SNR}[SNR]{signal-to-noise ratio}
\acrodef{FH}[FH]{frequency hopping}
\acrodef{BS}[BS]{base station}
\acrodef{UE}[UE]{user equipment}
\acrodef{UPA}[UPA]{uniform planar square array}
\acrodef{PPP}[PPP]{Poisson point process}
\acrodefplural{PPP}{Poisson point processes}
\acrodef{AWGN}[AWGN]{additive white Gaussian noise}
\acrodef{CDF}[CDF]{cumulative distribution function}
\acrodef{CCDF}[CCDF]{complementary cumulative distribution function}
\acrodef{ASE}[ASE]{area spectral efficiency}
\acrodef{MCS}[MCS]{modulation and coding scheme}
\acrodef{UCP}[UCP]{uniform clustering process}
\acrodef{LTE}[LTE]{long term evolution}

\begin{document}
\abovedisplayskip=0.5pt
\belowdisplayskip=0.5pt
\title
{Frequency Hopping on a \\ 5G Millimeter-Wave Uplink}
\author{\IEEEauthorblockN{Salvatore Talarico,
and Matthew C. Valenti} \IEEEauthorblockA
{West Virginia University, Morgantown, WV, USA} }%
\maketitle

\begin{abstract}
In order to overcome the anticipated tremendous growth in the volume of mobile data traffic, the next generation of cellular networks will need to exploit the large bandwidth offered by the millimeter-wave (mmWave) band.  A key distinguishing characteristic of mmWave is its use of highly directional and steerable antennas.  In addition, future networks will be highly densified through the proliferation of base stations and their supporting infrastructure.  With the aim of further improving the overall throughput of the network by mitigating the effect of frequency-selective fading and co-channel interference, 5G cellular networks are also expected to aggressively use frequency-hopping. This paper outlines an analytical framework that captures the main characteristics of a 5G cellular uplink. This framework is used to emphasize the benefits of network infrastructure densification, antenna directivity, mmWave propagation characteristics, and frequency hopping.
\end{abstract}

\section{Introduction}

Next generation (5G) cellular networks will require new technologies and frequency bands to meet the explosive growth in mobile data traffic \cite{Andrews:2014ccc}.  Continued use of existing microwave bands will not be able to meet these demands, and new networks must consider the use of millimeter wave (mmWave) technology \cite{Rangan:2014,Niu:2015}. Despite the common believe that mmWave might not be feasible for cellular networks due large near-field loss and poor penetration through obstacles, conducted measurements have shown the contrary \cite{Rappaport:2013}. In particular, it has been shown that mmWave cellular communications are feasible for cells with radiii on the order of 150-200 meters in densely deployed networks, provided that they are supported by a sufficient beamforming gain between the \acp{BS} and the mobiles that they serve \cite{Akdeniz:2014}.

Motivated by these encouraging results, several researchers have investigated the performance of mmWave cellular communications by  developing models that are able to capture the main characteristics, yet are analytical tractable. In \cite{Bai:2014}, the authors have proposed a framework for analyzing the coverage and rate in downlink mmWave cellular networks using tools from stochastic geometry. In particular, the authors account for blockages caused by buildings by using a distance-dependent \ac{LOS} probability function, and model the \acp{BS} as independent inhomogeneous \ac{LOS} and \ac{NLOS} \acp{PPP}. In \cite{DiRenzo:2015}, a framework for the analysis of the downlink of mmWave cellular networks is introduced, which incorporates path-loss and blockage models derived from reported experimental data. In \cite{Turgut:2015}, the authors analyze the average symbol error rate probability for the downlink of a mmWave cellular network. In \cite{Singh:2015}, the authors have proposed an analytical framework to characterize the rate distribution for both the downlink and the uplink in mmWave cellular networks by also accounting for self-backhauling \acp{BS}.

Another important characteristic of 5G networks is that they will likely maintain the basic structure of \ac{SC-FDMA} uplink systems, and support frequency-hopping as done to a certain extent in 4G systems. In order to account for these characteristics, \cite{torrieri:2015b} proposes a quasi-analytical approach to study a frequency-hopping mmWave cellular uplink. The current manuscript studies a similar system, but adopts a more theoretical approach that is more amenable to the tools of stochastic geometry.  The model accounts for millimeter-wave propagation, directional antenna beams, and frequency hopping. The analysis is used to highlight the benefits of antenna directivity, densification of the \acp{BS}, and frequency hopping.

The remainder of this paper is organized as follows. Sec. \ref{sec:Model} describes the network model, including the network topology, the propagation and antenna model, and the power control model. Sec. \ref{sec:Conditional_outage} derives a closed-form expression for the outage probability conditioned on a specific topology. Sec. \ref{sec:Average_outage} continues by describing a two step procedure for removing the conditioning to obtain a spatially averaged outage probability, wherein the first step is to decondition over the locations of the mobile devices and the second step is to decondition over the locations of the \acp{BS}.
Using this framework, Sec. \ref{sec:Numerical_Analysis} provides an evaluation of the performance of a typical mmWave cellular uplink. Finally, the paper concludes in Sec. \ref{sec:Conclusions}.

\section{Network Model}\label{sec:Model}

\begin{figure}[t!]
\centering
\hspace{-0.5cm}
\includegraphics[width=9.25cm]{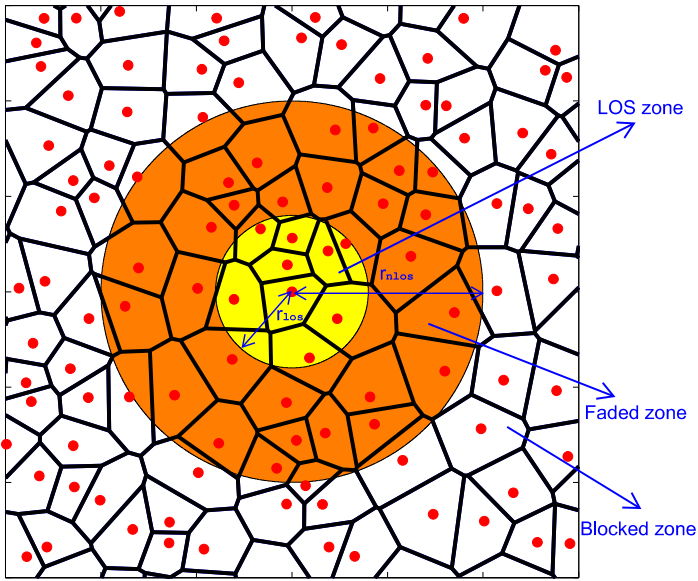}
\vspace{-0.7cm}
\caption{ Illustration of a typical network. The \acp{BS} are represented by filled red dots, while the tick black lines represents the Voronoi cells. The network is composed of three zones: a) \ac{LOS} zone, which is the inner circle; b) NLOS or faded zone, which is the annular region colored in orange; c) Blocked zone, which is the region outside the colored circle. \label{fig:zones} }
\vspace{-0.5cm}
\end{figure}

A typical network is shown in Fig. \ref{fig:zones}. As  depicted by the red dots in the figure, the network contains several \acp{BS}, each surrounded by a Voronoi cell.  The $j^{th}$ \ac{BS} and its location are represented by $Y_j$.  The reference \ac{BS} is $Y_0$, and the coordinate system is selected so that it is at the origin; i.e., $Y_0 = 0$.  The other \acp{BS} can placed in an arbitrary way.  For instance, they can be positioned according to the known locations of an actual network, or their locations may be synthesized from a stochastic-geometry model.  In Section IV, when we must specify the distribution of the \acp{BS}, we assume that they are drawn from a \textit{\ac{UCP}} \cite{torrieri:2013} with intensity $\lambda_\text{bs}$, where each \ac{BS} is surrounded by an \emph{exclusion zone} of radius $r_\text{min}$.

At mmWave frequencies, blockages by various obstacles often prevent \ac{LOS} propagation, but reflections may allow significant \ac{NLOS} propagation.  In order to model this phenomenon, the propagation model features distance-dependent models for path loss and fading.  The local-mean power, which is the average received power at a particular location in the network, is a function of the distance $d$ of the corresponding link. The path-loss function is expressed as the attenuation power law
\begin{equation}
f(d)  =  \frac{d^{-\alpha(d)}}{d_0^{-\alpha(d_0)}} , d \geq d_0
\end{equation}
where $\alpha(d)$ is the path-loss exponent, and $d_0$ is a reference distance that is less than or equal to the minimum of the near-field radius (typically assumed to be line-of-sight).

As shown in Fig. \ref{fig:zones}, the network is divided into three zones: a) the \textit{\ac{LOS} zone}; b) the \textit{faded zone}; c) and the \textit{blocked zone}.
The \ac{LOS} zone is a circular region $\mathcal A_\text{los}$ centered at the reference \ac{BS} with radius $r_\mathsf{los}$ and area $|\mathcal A_1|=\pi r_\mathsf{los}^2$. An interfering mobile in this region is assumed to be in the \ac{LOS} of the reference \ac{BS}, and its signal is assumed to be subject to \ac{AWGN} and has path loss $\alpha_\text{los} = 2$. An interferer within a distance $r_\mathsf{los}\leq d \leq r_\mathsf{nlos}$ from the reference \ac{BS} is in the \ac{NLOS} or \emph{faded} zone $\mathcal A_\text{nlos}$. While the signal from a mobile within it might still be harmful and received at the reference \ac{BS}, it is subject to Rayleigh fading and the path-loss exponent is $\alpha_\text{nlos} = 4$. An interferer that is beyond distance $r_\mathsf{nlos}$ is instead completely blocked, and therefore will not cause any interference.


Each transmitter and receiver within the network is equipped with an antenna array with highly directional beams in order to overcome the high propagation losses and power limitations of mmWave frequencies. Each \ac{BS} uses fixed sector beams to divide its coverage area into a fixed number of sectors centered at the \ac{BS}. Each mobile adaptively steers its beam in order to always face its serving \ac{BS}. At mmWave frequencies, sharp beams can be formed by using many antenna elements. A \ac{UPA} with  half-wavelength  antenna  element spacing \cite[Fig.1(b)]{Bai:2014} is assumed to be used at each transmitter and receiver. Let $\theta_\text{bs}$ and $\theta_\text{ms}$ indicate  the beamwidth of the antenna main-lobe in the azimuth of the \acp{BS} and mobiles, respectively. The antenna gain of the \acp{BS} is indicated by $a_\text{bs}$, while the antenna gain of the mobiles is $a_\text{ms}$. $N_\text{bs}$ is the number of antenna elements at each \ac{BS}, and $N_\text{ms}$ is the number of antenna elements equipped at each mobile. The antenna pattern is modeled by a two-level function with maximum gain equal to $G$ over the main-lobe and minimum gain $g$ elsewhere. The values for the beamwidth, main-lobe and back-lobe gains depend on the antenna elements $N$ of the \ac{UPA}, and can be evaluated per Table \ref{UPA_values}.

\begin{table}[t]
\caption{Antenna parameters of a uniform planar square antenna.}
\centering
\vspace{-0.25cm}
  \begin{tabular}{|c|c|}
    \hline
     Number of antenna elements& $N$\\
  \hline
  \hline
     Beamwidth $\theta$ & $\displaystyle \frac{2 \pi }{\sqrt{N}}$\\
  \hline
  \hline
     Main-lobe gain $G$ & $N$\\
  \hline
  \hline
     Side-lobe gain $g$ & $\displaystyle \frac{1}{\sin^2\left( \frac{3 \pi}{2\sqrt{N}}\right)}$\\
  \hline
  \end{tabular}
    \label{UPA_values}
    \vspace{-0.5cm}
\end{table}

Let $X_i = r_i e^{j \phi_i}$ represent the location of the $i^{th}$ mobile device.  For convenience, we represent $X_i$ as a complex number, so that $r_i=|X_i|$ is the distance to the origin (reference \ac{BS}) and $\phi_i$ is the azimuth angle from the \ac{BS} to the mobile.  The gain of the reference sector's receive antenna in the direction of $X_i$ is given by
\begin{equation}
a_i  = \left\{
\begin{array}
[c]{cc}%
G_\text{bs}, & \mbox{if $| \phi_i  - \psi | \leq \frac{\theta_{\text{bs}}}{2}$} \\
g_\text{bs}, & \text{otherwise}
\end{array}
\right.
\end{equation}
where $\psi$ is the offset angle of the beam pattern (i.e., the direction the antenna is pointing).  Similarly, the transmit antenna gain of $X_i$ in the direction of the reference receiver is
\begin{equation}
b_{i}
=
\left\{
\begin{array}
[c]{cc}%
G_\text{ms}, & \mbox{if $\left| \phi_i - \angle (Y_{g(i)}-X_i) \right| \leq \frac{\theta_{\text{ms}}}{2}$}\\
g_\text{ms}, & \text{otherwise}
\end{array}
\right.
\end{equation}
where $g(i)$ denotes a function that returns the index of the \ac{BS} serving $X_i$, and $\angle (Y_{g(i)}-X_i)$ is the angle from the \ac{BS} serving $X_i$ to $X_i$. By assuming that the \ac{BS} $Y_{g(i)}$ that serves mobile $X_i$ is the one with minimum local-mean path loss when the main-lobe of the transmit beam of $X_i$ is aligned with the sector beam of $Y_{g(i)}$, the serving \ac{BS} has index
\begin{eqnarray}
   g(i)
   & = &
\argmax_j { \left\{ f\left( |Y_j - X_i| \right) \right\}}.
\end{eqnarray}

A frequency-hopping \cite{torrieri:2015} \ac{SC-FDMA} uplink system is considered. Assuming the mobiles properly advance their signal transmissions, synchronous orthogonal frequency-hopping patterns can be allocated so that at any given instant in time, there is no intra-sector interference. The frequency-hopping patterns transmitted by mobiles in other cells (or sectors) are not generally orthogonal to the patterns in the reference sector, and hence produce inter-cell interference. In this paper, for the sake of tractability, it is assumed that all cells are synchronous and propagation delays are neglected.  Thus, when an inter-cell collision occurs (i.e., the mobiles in two different cells select the same hopping frequency), the collision persists for the entire slot; partial collisions are not considered here.

While a reference mobile is located within the reference sector, the potential interfering mobiles are drawn from a \ac{PPP} $\Phi$, which has intensity $\lambda_\text{ms}$ over the entire network except within the reference cell sector area $\mathcal A_{\text{sec}}$, where the intensity is zero. Mobiles are randomly re-located each time there is a frequency
hop. This assumes a large number of hops/users, and ignores the possibility that the same interferer could be colliding with the reference link multiple times during the same codeword.


The mobile hops $L$ times per codeword.  Assuming that full power control is used, the \ac{SINR} for a given codeword is
\begin{eqnarray}
   \gamma
   & = &
   \frac{ \sum_{t=1}^L h_{0,t} }
   {\mathsf{SNR}^{-1} + \sum_{t=1}^L I_t }
\end{eqnarray}
where
$\displaystyle \mathsf{SNR}$
is the \ac{SNR}
and
$h_{0,t}$ is the fading gain of the reference link during the $t^{th}$ hop, which is assumed to be gamma distributed with a large Nakagami-m factor $m_\text{los}$ as it is assumed that the reference link is \ac{LOS}. $I_t$ is the contribution from the interference during the $t^{th}$ hop. By separating the \ac{LOS} and \ac{NLOS} interfering contribution, it can be written as
\begin{eqnarray}
   I_t
   & = &
   \sum_{i \in \Phi_\mathsf{los}^t}
      \Omega_i
   +
   \sum_{i \in \Phi_\mathsf{nlos}^t}
       h_{i,t} \Omega_i \label{eq:Interference}
\end{eqnarray}
where $\Phi_\mathsf{los}^t$ is the set of LOS interferers during hop $t$, $\Phi_\mathsf{nlos}^t$ is the set of NLOS interferers during hop $t$. $h_{i,t}$ is the Rayleigh fading power of the $i^{th}$ NLOS interferer during the $t^{th}$ hop, which is modeled as an exponential random variable. $\Omega_i$ is the normalized power of the $i^{th}$ transmitter received at the reference \ac{BS}, given by
\begin{eqnarray}
   \Omega_i
    & =  & \frac{1}{L}
    \frac{a_i b_i}{G_\text{bs}G_\text{ms}}  \frac{f\left(|X_i|\right)}{f\left(|Y_{g(i)}-X_i|\right)}.
    \label{eq:Normalized received power}
\end{eqnarray}


\section{Conditional Outage Probability} \label{sec:Conditional_outage}

Let $\beta$ denote the minimum \ac{SINR} required for reliable reception of a codeword.
An \textit{outage} occurs when the \ac{SINR} falls below $\beta$.
The value of $\beta$ is related to the maximum achievable code rate $R$ that can be achieved.
Conditioning over the location of the interferers and \acp{BS} (which is embodied by the set $\mathbf \Omega = \{\Omega_i\}$), the outage probability averaged over the fading and frequency hopping is
\begin{eqnarray}
   p_{\mathsf{o}|\bf  \Omega}
   & = &
   \mathbb P\left[ \gamma < \beta\right | \bf \Omega ]
\end{eqnarray}
Define
\begin{eqnarray}
  h
  & = &
  \sum_{t=1}^L h_{0,t}
\end{eqnarray}

Since $h_{0,t}$ is gamma distributed with parameter $m_\text{los}$, $h$ is still gamma distributed but with parameter $M=m_\text{los}L$. Therefore, the outage probability can then be expressed as
\begin{eqnarray}
   p_{\mathsf{o}|\bf \Omega}
   & = &
   \mathbb P\left[ h < \beta\left( \mathsf{SNR}^{-1} + \sum_{t=1}^L I_t \right) \right] \nonumber \\
   &=&
   \mathbb E_{\bf h} \left[ F_h \left( \beta \left( \mathsf{SNR}^{-1} + \sum_{t=1}^L I_t \right) \right) \right] \label{eq:Outage_2}
\end{eqnarray}
where $\mathbb E_{\bf h} \left[\cdot \right]$ is the average over the set of fading gains $\{ h_{it}\}$ for the \ac{NLOS} interferers, and  $F_h \left(\cdot \right)$ is the \ac{CDF} of $h$. For a normalized gamma distributed random variable $Z$ with (integer) parameter $m$, the \ac{CDF} evaluated at a point $z$ can be tightly lower bounded \cite{Alzer:1997} as follows
\begin{eqnarray}
  F_Z(z)
  & = &
  \left( 1-\exp\left\{ -m (m!)^{\frac{-1}{m}} z \right\} \right)^m \label{eq:CDF_h}
\end{eqnarray}
Substituting (\ref{eq:CDF_h}) into (\ref{eq:Outage_2}) and defining
$\tilde{M} = {(M!)^{\frac{-1}{M}}}$ yields
\begin{eqnarray}
   p_{\mathsf{o}|\bf \Omega}
   & = & \mathbb E_{\bf h}\left[
   \left( 1-e^{-M \tilde{M} \beta \left( \mathsf{SNR}^{-1} + \sum_{t=1}^L I_t \right) } \right)^M\right] \label{eq:Outage_3}
\end{eqnarray}
Substituting (\ref{eq:Interference}) in (\ref{eq:Outage_3}) and performing a binomial expansion,
\begin{eqnarray}
   p_{\mathsf{o}| \bf \Omega}
   & = &
   1-\sum_{l=1}^{M} {{M}\choose{l}} \left( -1 \right)^{l+1} \times \nonumber \\
   & & \exp{\left[ - l M \tilde{M} \beta \left( \mathsf{SNR}^{-1} + \sum_{t=1}^L \sum_{i \in \Phi_\mathsf{LOS}^t}
      \Omega_i \right)\right] } \times \nonumber
      \\ & & E_{\bold h}\left[ \exp{\left( - l M \tilde{M} \beta \sum_{t=1}^{M}\sum_{i \in \Phi_\mathsf{NLOS}^t}
       h_{i,t} \Omega_i \right)}\right]. \label{eq:outage_1}
\end{eqnarray}

Using the fact that the set of fading gains $\{h_{i,t}\}$ are independently and identically distributed (iid), then (\ref{eq:outage_1}) can be written as follows
\begin{eqnarray}
   p_{\mathsf{o}|\bf \Omega}
   & = &
   1-\sum_{l=1}^{M} {{M}\choose{l}} \left( -1 \right)^{l+1} \times \nonumber \\
   & & \exp{\left[ - l M \tilde{M} \beta \left( \mathsf{SNR}^{-1} + \sum_{t=1}^L \sum_{i \in \Phi_\mathsf{LOS}^t}
      \Omega_i \right)\right] } \times \nonumber
      \\ & & \prod_{t=1}^{L} \prod_{i \in \Phi_\mathsf{NLOS}^t} E_{h_{i,t}}\left[ \exp{\left( - l M \tilde{M} \beta
       h_{i,t} \Omega_i \right)}\right]. \label{eq:outage_2}
\end{eqnarray}

Solving the expectation in (\ref{eq:outage_2}) for the gamma-distributed $h_{i,t}$ with parameter $m_\text{nlos}$ yields
\begin{eqnarray}
   p_{\mathsf{o}|\bf \Omega}
   & = &
   1-\sum_{l=1}^{M} {{M}\choose{l}} \left( -1 \right)^{l+1} \times
   \nonumber \\
   &  &\exp{\left[ - l M \tilde{M} \beta \left( \mathsf{SNR}^{-1} + \sum_{t=1}^L \sum_{i \in \Phi_\mathsf{LOS}^t}
      \Omega_i \right)\right] } \times \nonumber
      \\ & & \prod_{t=1}^{L} \prod_{i \in \Phi_\mathsf{NLOS}^t} \left( 1+ \frac{l \tilde{M} M \beta \Omega_i}{m_\text{nlos}}\right)^{-m_\text{nlos}}. \label{eq:outage_3}
\end{eqnarray}


\begin{table}[t]
\vspace{0.25cm}
\caption{Settings used. }
\vspace{-0.25cm}
\centering
  \begin{tabular}{|c|c|}
    \hline
     Faded zone outer radius & $r_\text{nlos}=10$\\
  \hline
  \hline
     LOS zone radius & $r_\text{los}=2$\\
  \hline
  \hline
     Distribution for the \acp{BS} & \ac{UCP} with $r_\text{min}=1$ \\
  \hline
  \hline
         Distribution of the mobiles & PPP \\
  \hline
  \hline
       Intensity of \acp{BS} & $\lambda_\text{bs}=0.2$ \\
  \hline
  \hline
       Intensity of mobiles & $\lambda_\text{ms}=1$ \\
  \hline
  \hline
    Path loss exponent  & $\alpha_\text{los}=2$ and $\alpha_\text{nlos}=4$\\
  \hline
  \hline
   Fading parameters &    $m_\text{los}=3$ and $m_\text{nlos}$=1     \\
  \hline
  \hline
  Reference distance    &   $d_{0}=0.01$ \\
  \hline
  \hline
  Signal-to-noise ratio &   $ \displaystyle \mathsf{SNR} = 25$ dB \\
  \hline
  \hline
  Antenna elements at the \ac{BS}  &   $N_\text{bs}=256$ \\
  \hline
  \hline
  Antenna elements at the mobile  &   $N_\text{ms}=16$ \\
  \hline
  \hline
 Number of frequency hops   &   $L=2$ \\
  \hline
  \hline
  Outage contraint   &   $\hat \epsilon=0.1$ \\
  \hline
  \hline
  Network trials & $N_\text{trials}=10^6$  \\
  \hline
  \end{tabular}
    \label{main_table}
    \vspace{-0.5cm}
\end{table}

\section{Spatially Averaged Outage Probability} \label{sec:Average_outage}

Typical analyses based on stochastic geometry find the outage probability averaged over the network topology.  Since the locations of the mobile stations change more frequently than the locations of the \acp{BS}, a sensible approach is to find the spatially averaged outage probability of the network by first averaging over the locations of the mobiles, then averaging over the locations of the \acp{BS}.  As the locations of the mobiles are contained in the point process $\Phi$, the outage probability conditioned on only the locations of the \acp{BS}, but averaged over the locations of the mobiles, can be written as
\begin{eqnarray}
   p_{\mathsf{o}|{\bf  Y} }
   & = &
   \mathbb P\left[ \gamma < \beta\right | {\bf Y} ] = \mathbb E_\Phi \left[ p_{\mathsf{o}|\bf  \Omega }\right]. \label{eq:unconditioned_OP2}
\end{eqnarray}
where $\mathbf Y$ is the set of \ac{BS} locations.

Substituting (\ref{eq:outage_3}) into (\ref{eq:unconditioned_OP2}) yields
\begin{eqnarray}
   p_{\mathsf{o}|\bf  Y}
   & = &
   1-\sum_{l=1}^{M} {{M}\choose{l}} \left( -1 \right)^{l+1} \times
\nonumber \\
& & \exp{\left[ - l M \tilde{M} \beta \left( \mathsf{SNR}^{-1} + L \mathcal E_1 \right)\right] } \mathcal E_2^L \label{eq:outage_4}
\end{eqnarray}
where
\begin{eqnarray}
\mathcal E_1 &=&\mathbb E_\Phi \left[ \sum_{i \in \Phi_\mathsf{LOS}^t}
      \Omega_i \right] \label{eq:average1}\nonumber \\
      &\stackrel{a}{=}& \lambda_{mb} \left(|\mathcal A_1|-|\mathcal A_{\text{sec}}|\right) \mathbb E_X\left[ \Omega_i \right] \nonumber \\
\mathcal E_2 &=&  \mathbb E_\Phi \left[ \prod_{i \in \Phi_\mathsf{NLOS}^t} \left( 1+ \frac{l \tilde{M} M \beta \Omega_i}{m_\text{nlos}}\right)^{-m_\text{nlos}}\right] \nonumber \\
&\stackrel{b}{=}& \exp\left\{ \hspace{-0.05cm} - \hspace{-0.05cm} \lambda_{mb} |\mathcal A_2| \hspace{-0.05cm} \left( \hspace{-0.05cm} 1 \hspace{-0.05cm} - \hspace{-0.05cm} \mathbb E_X \hspace{-0.15cm} \left[ \hspace{-0.05cm} \left( \hspace{-0.05cm} 1 \hspace{-0.05cm} + \hspace{-0.05cm} \frac{l \tilde{M} M \beta \Omega_i}{m_\text{nlos}}\right)^{-m_\text{nlos}} \right] \hspace{-0.05cm} \right)\hspace{-0.15cm}\right\} \nonumber
\end{eqnarray}
where $(\stackrel{a}{=})$ and $(\stackrel{b}{=})$ are obtained using a similar approach of that taken in \cite{valenti:2014} by using the fact that the mobiles are independently and uniformly distributed, and the number of mobiles within an area is a Poisson random variable. For both $\mathcal E_1$ and $\mathcal E_2$, the final expression reduces to the evaluation of $E_X\left[ \cdot\right]$, which is the average with respect to the location $X$ of a single mobile. These two averages can be obtained either by Monte Carlo simulation (by simulating the location of a mobile within $\mathcal A_1 \backslash \mathcal A_{\text{sec}}$ and $\mathcal A_2$, respectively) or by using numerical integration after the distribution of the \ac{CDF} of the normalized received power is evaluated using the approach provided in \cite{Pure:2015}.

Having averaged the outage probability over the locations of the mobiles, the outage probability averaged over the locations of the \acp{BS} can be found, if so desired, by averaging (\ref{eq:unconditioned_OP2}) over the spatial distribution of the \acp{BS}.  The average outage probability is hence
\begin{eqnarray}
   p_{\mathsf{o}}
   & = &
   \mathbb P\left[ \gamma < \beta\right] = \mathbb E_Y \left[ p_{\mathsf{o}|\bf  Y }\right].
\end{eqnarray}
In order to evaluate $p_{\mathsf{o}}$, a Monte Carlo approach can be used. After a large set of network topologies, say $N_\text{trial}$ network realizations,
 is created, (\ref{eq:outage_4}) is evaluated and saved for each realization, and these values are used to evaluate $p_{\mathsf{o}}$ by computing their average.  Arguably, a more useful quantity that the overall average outage probability is the distribution of $ p_{\mathsf{o}|\bf  \Omega}$, which gives insight into the number of network realizations (or the number of reference sectors in a large network) that achieve a target outage probability; such a distribution can also be found using Monte Carlo methods, as shown in the next section.

\begin{figure}[t]
\centering
\includegraphics[width=8.75cm]{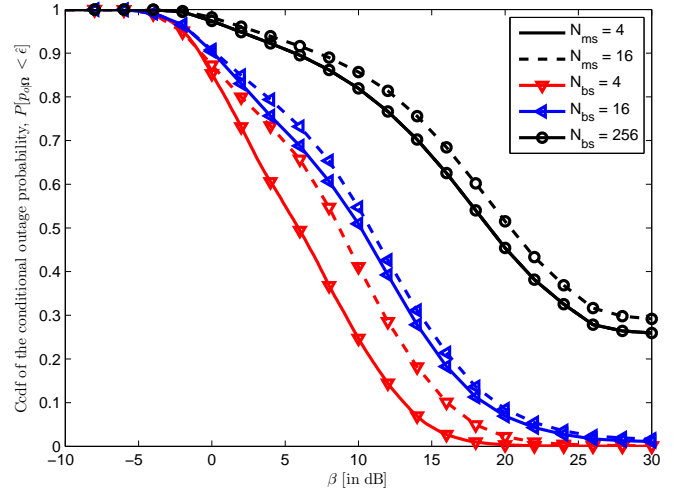}
\vspace{-0.7cm}
\caption{ \ac{CCDF} of the conditional outage probability $p_{\mathsf{o}|\bf  \Omega}$ as function of the \ac{SINR} threshold. Curves are parameterized over the antenna elements equipped at both the \ac{BS} and mobile. \label{Figure:AP_CCDF_OP} }
\vspace{-0.5cm}
\end{figure}

\section{Numerical Results} \label{sec:Numerical_Analysis}

This section evaluates the performance of a 5G mmWave uplink by computing the \ac{CCDF} of the conditional outage probability $p_{\mathsf{o}|\bf  \Omega}$. In particular, the \ac{CCDF} of $p_{\mathsf{o}|\bf  \Omega}$ is used to depict the percentage of the mobiles that have an outage probability below a given outage constraint $\hat \epsilon$, i.e $\mathbb P\left[ p_{\mathsf{o}|\bf  \Omega} < \hat\epsilon \right]$. While $p_{\mathsf{o}|\bf  \Omega}$ can be evaluated analytically using (\ref{eq:outage_3}), its distribution is computed using a Monte Carlo approach by randomly varying at each of the $N_\text{trial}$ trials the topology of the network according to the spatial distribution of both the mobiles and the \acp{BS}.

Another useful metric is the \textit{\ac{ASE}}. This metric provides the maximum rate of successful data transmissions per unit area, and it is defined as
\begin{eqnarray}
   \tau_\text{ASE}
   & = &
   \lambda_\text{bs} R \left( 1- p_{\mathsf{o}} \right)
\end{eqnarray}
where the units are bits per channel use (bpcu) per unit area. If a capacity approaching code is assumed to be used and rate adaptation is performed over a large set of \acp{MCS}, the code rate $R$ can be related to the \ac{SINR} threshold using the Shannon bound for complex discrete-time \ac{AWGN} channels, as follows
\begin{eqnarray}
   R
   & = & \log_2\left( 1+ l_s \beta\right)
\end{eqnarray}
where $l_s=0.794$ to account for the typical $1$ dB loss from Shannon capacity of an actual code.

In the following example, the settings summarized in Table \ref{main_table} are adopted, if not otherwise stated. Fig. \ref{Figure:AP_CCDF_OP} shows the \ac{CCDF} of the conditional outage probability $p_{\mathsf{o}|\bf  \Omega}$ as function of the \ac{SINR} threshold when the \ac{BS} and the mobile are equipped with a different number of antenna elements. This figure highlights the importance of sectorization, and that a small improvement is gained by equipping large arrays of antennas at the mobile.

\begin{figure}[t]
\centering
\includegraphics[width=8.75cm]{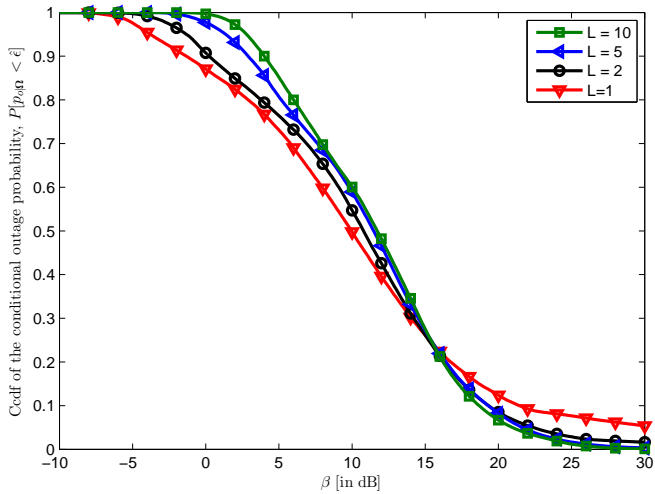}
\vspace{-0.7cm}
\caption{ \ac{CCDF} of the conditional outage probability $p_{\mathsf{o}|\bf  \Omega}$ as function of the \ac{SINR} threshold. Curves are parameterized over the number of frequency hops per codeword. \label{Figure:NH_CCDF_OP} }
\end{figure}

Fig. \ref{Figure:NH_CCDF_OP} compares the performance of the network in terms of the number of hops per codeword, again by showing the \ac{CCDF} of the conditional outage probability $p_{\mathsf{o}|\bf  \Omega}$ as function of the \ac{SINR} threshold. This figure shows that the percentage of mobiles
that are able to satisfy an outage constraint $\hat \epsilon$ sensibly increases as the number of hops per codeword is increased emphasizing the role of frequency-hopping in reducing the power level of the inter-cell interferers.

\begin{figure}[t]
\centering
\includegraphics[width=8.75cm]{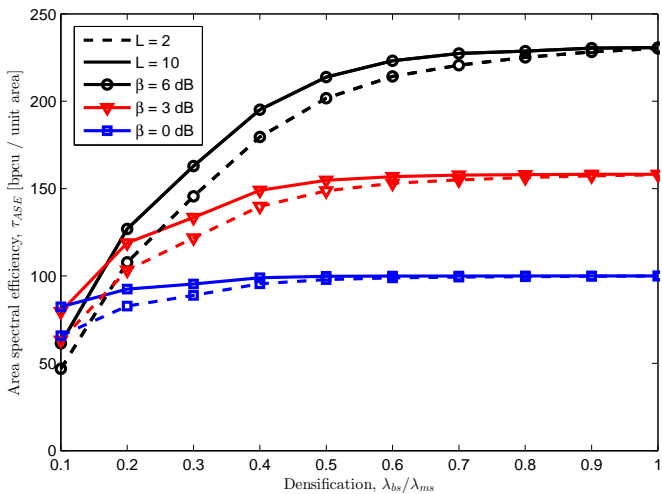}
\vspace{-0.7cm}
\caption{ \ac{ASE} as function of densification of the \acp{BS}. Curves are parameterized by both the \ac{SINR} threshold and the number of frequency hops per codeword. \label{Figure:AOP} }
\vspace{-0.5cm}
\end{figure}

Fig. \ref{Figure:AOP} provides the \ac{ASE} as the network densifies by deploying a larger number of \acp{BS}. This figure shows the importance of densification. Increases in the \ac{SINR} threshold $\beta$ degrade the outage probability. However, for a sufficiently densified network this effect is minor compared with increased code rate that can be accommodated. As a result, the area spectral efficiency $\tau_\text{ASE}$ increases significantly. Furthermore, as the network becomes more densified a lower power is required for each intended link due to power control, and therefore the power level of the interference decreases with the consequence that aggressive hopping doesn't provide any significant improvement.

\section{Conclusions} \label{sec:Conclusions}

This paper derives an analytical framework to evaluate the outage probability for mmWave uplink cellular networks when frequency-hopping is adopted. The model includes the effects of mmWave propagation, highly directional beams, and arbitrary network topologies. Results emphasize the importance of \ac{BS} densification, and sectorization. Furthermore, the beneficial effects of frequency-hopping are illuminated by the analysis, and frequency hopping seems to be a good complement to the use of power control and directional antennas.

\bibliographystyle{ieeetr}
\bibliography{AsilomarRefs}

\balance
\vfill

\end{document}